\documentclass[fleqn,twoside]{article}
\usepackage{espcrc2}

\usepackage{graphicx}
\usepackage[figuresright]{rotating}

\newcommand{\AmS}{{\protect\the\textfont2
  A\kern-.1667em\lower.5ex\hbox{M}\kern-.125emS}}

\title{Excited States in Staggered Meson Propagators}

\author{The MILC Collaboration: C. Bernard\address{Department of Physics, Washington University,
                           St.~Louis, MO 63130, USA \vskip -0.1cm },
       T. Burch \address{Institut f\"ur Theoretische Physik, 
                        Universit\"at Regensburg, D-93040 Regensburg, 
                        Germany\vskip -0.1cm},
       C. DeTar \address[UTAH]{Physics Department, University of Utah, 
                           Salt Lake City, UT 84112, USA\vskip -0.1cm},
       Steven Gottlieb \address{Department of Physics, Indiana University,
                           Bloomington, IN 47405, USA\vskip -0.1cm},
       E.B. Gregory \address[ARIZ]{Department of Physics, 
                           University of Arizona, Tucson, AZ 85721, 
                           USA\vskip -0.1cm}\thanks{Presented by E.B. Gregory},
       U.M. Heller \address{American Physical Society, One Research Road, 
                           Box 9000, Ridge, NY 11961, USA\vskip -0.1cm },
       J. Osborn \addressmark[UTAH],
       R. Sugar \address{Department of Physics, University of California, 
                           Santa Barbara, CA 93106, USA\vskip -.1cm},
       D. Toussaint \addressmark[ARIZ]}
       
\begin{document}

\begin{abstract}
We report on preliminary results from multi-particle fits to 
meson propagators with three flavors of light dynamical quarks. 
We are able to measure excited states in propagators with pion quantum numbers,
which we interpret as the pion $2S$ state, and argue is evidence of 
locality of the action.
In the $a_0$ ($0^{++}$) propagators we find evidence for excited
states which are probably the expected decay channels,
$\pi+\eta$ and $K+\overline{K}$.
\vspace{-0.25in}
\end{abstract}

\maketitle
\section{INTRODUCTION}
Extracting ground state hadron masses from lattice operator correlation 
functions has become a standard exercise, and results generally  agree 
with experimental 
measurements.
(See, for example \cite{Bernard:2001av,Bernard:2002bk,Allton:2001sk,Aoki:2002uc,Blum:2000kn,Aoki:2002fd}.)
However, determining excited state masses on the lattice
(e.g.\ \cite{Maynard:2002ys,Richards:2001bx,Lacock:1996vy,Fiebig:2001nn,Chen:2000qj})
 is more difficult.
Such determinations can provide tests of the particular lattice approach,
can be used to study decays of unstable hadrons and, in principle, could
help clarify quark model interpretations of higher mass hadrons.

\begin{table}[hb]
\vspace{-0.3in}
\label{sim_lattices}
\caption{}
\vspace{-0.1in}
\begin{tabular}{cllll}
$\beta=10/g^2$ & $am_q$, $am_s$ & $L^3\times T$& $n_{\rm config}$ \\
\hline
\hline
6.96 & 0.100, 0.100 & $20^3\times 64$ & 339 \\
6.85 & 0.050, 0.050 & $20^3\times 64$ & 414 \\
6.83 & 0.040, 0.050 & $20^3\times 64$ & 275 \\
6.81 & 0.030, 0.050 & $20^3\times 64$ & 564 \\
6.76 & 0.010, 0.050 & $20^3\times 64$ & 658 \\
6.76 & 0.007, 0.050 & $20^3\times 64$ & 443 \\
6.76 & 0.005, 0.050 & $20^3\times 64$ & 159 \\
\hline
7.18 & 0.0310, 0.031 & $28^3\times 96$ & 496 \\
7.11 & 0.0124, 0.031 & $28^3\times 96$ & 527 \\
7.09 & 0.0062, 0.031 & $28^3\times 96$ & 592 \\
\end{tabular}
\end{table}

\section{SIMULATION}

We measured meson propagators on lattices generated using 
the ``Asqtad'' action\cite{Asqtad} with $2+1$ flavors of dynamical quarks, 
that is, one flavor with mass approximately equal to the physical strange 
quark mass and two lighter flavors with  degenerate masses. 
(See Table \ref{sim_lattices}.)
We describe the $20^3\times 64$ lattices as ``coarse'' lattices with
lattice spacing of $a\approx 0.12$ fm, while the  $28^3\times 96$ ``fine''
lattices have a spacing of $a\approx 0.09$ fm.
More detailed descriptions of these simulations and analysis procedures can
be found in Refs.~\cite{Bernard:2001av,Bernard:2002bk}.
In this report we discuss excited states in the pseudoscalar ($0^{-+}$) and
scalar ($0^{++}$) isovector meson propagators.

\begin{figure}[htb]
\resizebox{2.8in}{2.8in}{\includegraphics{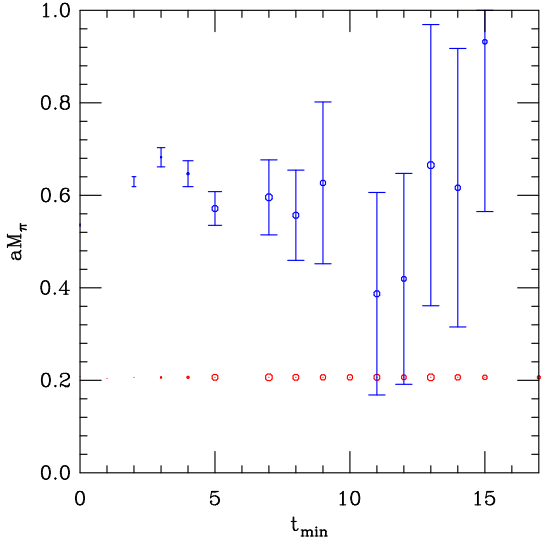}}
\vspace{-0.4in}
\caption{Pion fits for $10/g^2=7.11$, $t_{\rm max}=28$. 
$t_{\rm min}=5$ was selected.
Symbol size is proportionate to confidence level.}
\label{pion_fit}
\vspace{-0.2in}
\end{figure}

\section{RESULTS AND CONCLUSIONS}
\subsection{$0^{-+}$ states}
In the staggered fermion formulation propagators generally appear with an
oscillating component that represents a parity partner state. There is no 
partner to the taste-pseudoscalar pion and $s\overline{s}$ states,
hence these propagators contain
contributions only from $J^{PC}=0^{-+}$ states. 
Moreover, the signal to noise ratio is large at all time
separations, so it is relatively easy to measure the contribution of 
excited states.
\begin{figure}[htb]
\scalebox{0.45}{\includegraphics{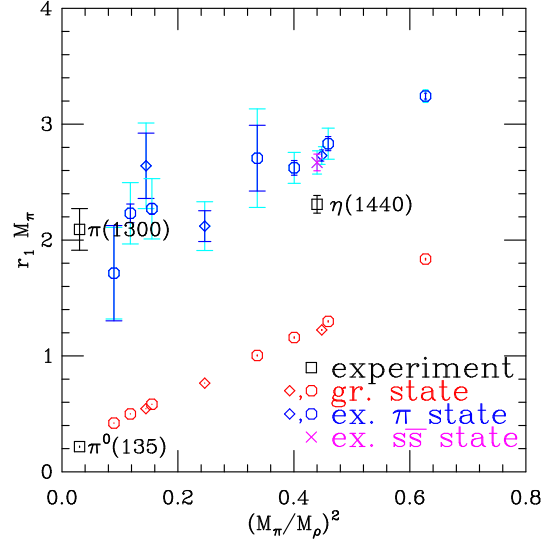}}
\vspace{-0.4in}
\caption{Ground state and excited state pion masses. Diamonds are fine 
lattice data; octagons are coarse lattice data. $s\overline{s}$ states are 
shown at $(M_{s\overline{s}}/M_{\phi})^2=(680/1020)^2=0.44$, where  
$M_{s\overline{s}}=\sqrt{ 2M_K^2 - M_\pi^2 }=680$MeV.  }
\label{excited_pions}
\vspace{-0.2in}
\end{figure}

We fit $0^{-+}$ propagators to the form:
\begin{eqnarray}
P(t)&=A_0(e^{-m_0t} + e^{-m_0(T-t)}) \\ \nonumber
&+A_1(e^{-m_1t} + e^{-m_1(T-t)}),
\end{eqnarray}
where $T$ is the length of the time dimension of the lattice, $t$ is the
time separation of the source and sink operators and $m_0$ and $m_1$ are
the masses of the ground state and lowest excited states.
We look for high-confidence, low-error fits that fall in plateaus of 
stability of the fit parameters with respect to the limits 
$t_{\rm max}$ and $t_{\rm min}$ of the propagator included in the fit.
Figure \ref{pion_fit} shows a set of pion fits with varying $t_{\rm min}$,
and $t_{\rm max}$ fixed at 28.

Figure \ref{excited_pions} shows the results of this fitting 
procedure for all of the data sets. The darker error bars are
statistical, and lighter error bars represent
systematic error due to fit choice. We plot the product of ground state and 
first excited state masses with $r_1$, a length scale derived from the 
static quark potential, against the ratio $(m_{\pi}/m_{\rho})^2$, a measure 
of the quark mass. At the physical value of $(m_{\pi}/m_{\rho})^2=0.03$ we
plot the experimental mass of the states $\pi^0(135)$ and the $\pi(1300)$ scaled 
with $r_1=0.317(7)$ fm\cite{Davies:2003ik}. 
Also, an excited 
$0^{-+}$ $s\overline{s}$ state, an average of measurements on all of the fine
 lattices, is compared to the experimental mass of the $\eta(1440)$, with
which agreement is very marginal.

We feel that the consistency of 
the simulation results with  the $\pi(1300)$ is an important indication that  
the fourth root of the determinant in the action does not introduce a  
non-locality with scale larger than the $\pi(1300)$ Compton wavelength: 0.15 fm.

\subsection{$0^{++}$ states}

\begin{figure}[htb]\vspace{-0.3in}
\resizebox{2.8in}{2.8in}{\includegraphics{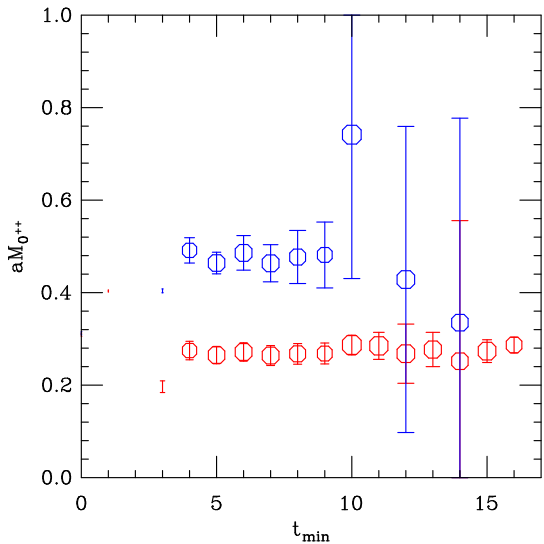}}
\vspace{-0.4in}
\caption{Fits of $0^{++}$ states for $10/g^2=7.09$ $t_{\rm max}=35$.
Symbol size is proportional to confidence level.
$t_{\rm min}=5$ was selected.}
\label{a0_masses}
\vspace{-0.4in}
\end{figure}

The $a_0$ meson ($J^{PC}=0^{++}$) propagator occurs as the oscillating 
parity partner 
in the taste-scalar pion propagator. With states of both parities contributing 
to the propagator, it is more difficult to resolve excited states.
We fit these $a_0-\pi$ propagators to the form:
\begin{eqnarray}
P(t)&=&A_0(e^{-m_0t} + e^{-m_0(T-t)})  \\ \nonumber
&+&A_1(e^{-m_1t} + e^{-m_1(T-t)})      \\ \nonumber
&+&A_2(-1)^t(e^{-m_2t} + e^{-m_2(T-t)})\\ \nonumber
&+&A_3(-1)^t(e^{-m_3t} + e^{-m_3(T-t)}).
\end{eqnarray}
Here $m_2$ and $m_3$ are the $0^{++}$ masses. In one case, 
$\beta=7.09$, we were able to unambiguously discern an excited
$0^{++}$ state. See Figure \ref{a0_masses}.  

\begin{figure}[htb]
\scalebox{0.45}{\includegraphics{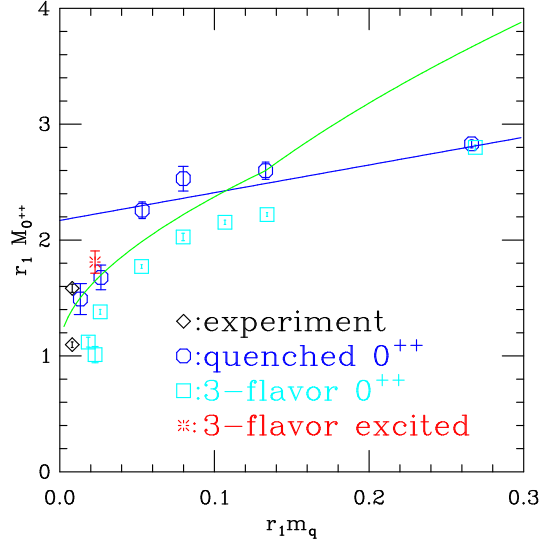}}
\vspace{-0.4in}
\caption{Mass of $0^{++}$ states as a function of $r_1m_q$, with 
$r_1=0.317(7)$ fm.
Straight line is a fit to mass $a_0$ with heavier quenched
quarks.  Curved line is the mass of $\pi+\eta$. 
Physical states are $\pi+\eta$ (682MeV) and 
$a_0$/$K\overline{K}$ (980MeV).}
\label{a0_fits}
\vspace{-0.3in}
\end{figure}

The $0^{++}$ states are interesting in that several of the lowest lying 
states are likely not pure $q\overline{q}$ states because the two
meson states to which the $a_0$ decays are accessible in our simulations.
In Figure \ref{a0_fits}, 
the straight line is an extrapolation of the heavier-$m_q$ quenched $a_0$ 
masses, and is an expectation of the mass of a $q\overline{q}$ state 
with $0^{++}$. The curved
line, which crosses below the ``$a_0$-line'' at low $m_q$ traces the sum of the $\pi$
and $\eta$ ground state masses\cite{Bernard:2001av}.  The fitted
 full-QCD $0^{++}$ ground state masses roughly trace the $\pi+\eta$ line,
possibly displaying the lower half of an avoided level crossing, an 
indication of the decay of the pure $q\overline{q}$ $a_0$. The measured
$0^{++}$ excited state (burst) is not consistent with the straight line 
either, but relatively close to the experimental value of the mass of a 
$K\overline{K}$ molecule. Future efforts will try to fill out 
Figure \ref{a0_fits} with more $0^{++}$ excited states.


\end{document}